\documentclass{webofc}
\usepackage[varg]{txfonts}   
\begin{document}
\title{Exploring the Spin Structure of the Nucleon at STAR}
%
%

\author{\firstname{Ting} \lastname{Lin}\inst{1}\fnsep\thanks{\email{tinglin@sdu.edu.cn}} for the STAR Collaboration
}

\institute{
Institute of Frontier and Interdisciplinary Science \& Key Laboratory of Particle Physics and Particle Irradiation (MoE), Shandong University, Qingdao, Shandong 266237, China
          }

\abstract{%

Understanding the internal spin structure of the nucleon remains a challenge in strong interaction physics. The unique capability of RHIC opened new avenues in studying the internal structure of the proton with unprecedented depth and precision. Significant progress has been made in the last few years through various measurements at STAR. The longitudinal spin measurements have contributed significantly to our understanding of the quark and gluon helicity distributions inside the proton. The longitudinal double-spin asymmetry, $A_{LL}$, from STAR inclusive jet and dijet measurements provides the first evidence of a positive gluon polarization with partonic momentum fraction $x > 0.05$. The reconstruction of $W^{\pm}/Z$ in longitudinally polarized proton-proton collisions indicates that there is a flavor separation of the light sea quark helicity distributions. In transversely polarized proton collisions, $W^{\pm}/Z$-bosons provide the first constraint on the sea-quark Sivers function and contributes to the tests of the predicted sign change. The tilt of the dijet opening angle provides a direct access to the first Mellin momentum of the Sivers function and avoids the spin-correlated fragmentation contributions. The novel measurements of the azimuthal distributions of identified hadrons in jets and spin-dependent dihadron correlations directly probe the collinear quark transversity in the proton, with the former coupled to the transverse momentum dependent (TMD) Collins fragmentation function and the latter to the dihadron interference fragmentation function. These measurements shed lights on Sivers function, quark transversity and spin-dependent fragmentation functions in both collinear and TMD formalism. In this proceeding, recent jet results on both the longitudinal and transverse spin structure of the proton from STAR are presented. 
}
\maketitle
\section{Introduction}
\label{intro}
Proton and neutron constitute the atomic nucleus, and hence all the visible mass in the universe. They also possess internal structures, that are constructed of three valence quarks and quark-antiquark pairs which are all bounded together by gluons. The strong interactions of these constituents, which still contain many unknowns, determine the observed properties of the nucleon, including masses, spins, magnetic moments, and their responses to external forces. As the next generation of nuclear physics facility --- Electron-Ion Collider (EIC) is approaching, there are growing interests to explore the internal structure of the nucleon, especially its spin structure~\cite{Accardi:2012qut, Anderle:2021wcy}.\par

The Relativistic Heavy Ion Collider (RHIC) is the first and only collider in the world that is capable of colliding both the longitudinally and transversely polarized proton beams up to center-of-mass energy of 510~GeV. It is a unique tool to explore the puzzle of the proton's spin physics. When combined with data from the future EIC, they will establish the limits of factorization and validate its universality, thus enabling a deeper understanding of fundamental quantum chromodynamics (QCD)~\cite{Alekseev:2003sk}.\par

The Solenoidal Tracker at RHIC (STAR) is one of the experiments located at RHIC facility, with over two decades of successful operation. The central part of the detector includes the electromagnetic calorimetry and charge particle tracking, enabling the analysis of various final state interactions within the experiment. In polarized $pp$ collisions, Barrel and Endcap electromagnetic calorimeters also served as triggering detectors, e.g. jet events, in addition to precision measurement of electromagnetic particles. Jet events are selected through the jet patch trigger algorithm if the energy depositions within $\Delta\eta\times\Delta\phi = 1\times1$ regions exceed a given threshold~\cite{STAR:2002eio}.\par

Jet is a collimated spray of particles stemming from the fragmentation and hadronization of quarks and gluons in hadronic interactions. It plays a crucial role in experimental particle physics, serving as a distinctive signature of underlying quark and gluon dynamics. STAR is a very good jet detector with its full coverage in $\phi$ and quite wide range in $\eta$ ($-1 < \eta < 2$ with Barrel and Endcap electromagnetic calorimeters). The analysis of jets, from either longitudinally or transversely polarized proton-proton collisions, offers valuable insights into the internal spin structure of the proton.\par

\section{Longitudinal Spin Structure}
\label{sec-1}

\begin{figure}[h]
\centering
\includegraphics[width=7.7cm,clip]{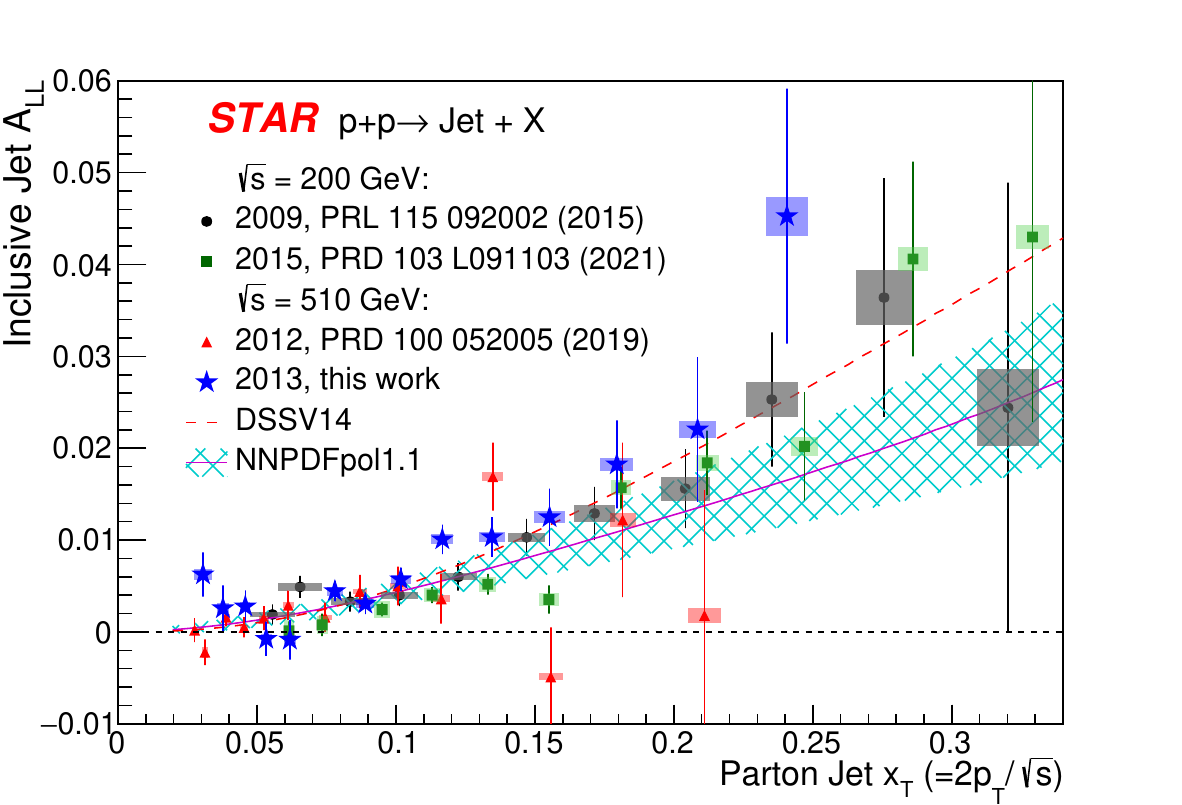}
\includegraphics[width=5cm,clip]{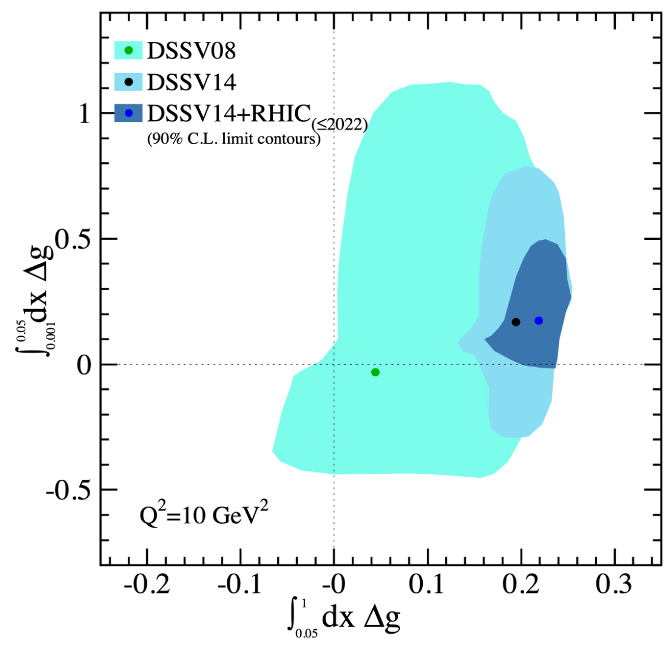}
\caption{Left panel: Inclusive jet $A_{LL}$ versus $x_{T}$ from~\cite{STAR:2021mqa}. Right panel: Integral of the gluon helicity from~\cite{RHICSPIN:2023zxx}.}
\label{fig-1}       
\end{figure}

Recent measurements of the inclusive jet $A_{LL}$ from the longitudinally polarized proton-proton collisions at STAR has significantly advanced our understanding of the proton's spin decomposition in canonical approach. In this framework, the proton's intrinsic spin of $\hbar/2$ is attributed to the contributions from quark spin, gluon spin, and their respective orbital angular momenta. The quark and gluon spin contributions are quantified through the integral of the associated spin-dependent parton distribution functions, commonly referred to as helicity. The observed $A_{LL}$, which measures the ratio of longitudinally polarized cross section over the unpolarized one, is proportional to the helicity of the two interacting partons~\cite{Jaffe:1989jz}.\par

At RHIC energy, gluon-gluon and quark-gluon hard scatterings are the dominant leading-order interactions for inclusive jet production, which make the $A_{LL}$ an ideal probe to the gluon polarization. The left panel of Fig.~\ref{fig-1} illustrates a decade's effort of the inclusive jet $A_{LL}$ results from STAR~\cite{STAR:2021mqa}. In this figure, $A_{LL}$ is plotted as a function of jet $x_{T}$, which is jet $p_{T}$ over the beam energy, thus we can compare measurements from different collision energies. Good consistency is observed among these measurements, reinforcing the reliability of the results. Furthermore, they also align well with theoretical expectations from DSSV and NNPDF groups, both indicate positive gluon polarization inside proton~\cite{Nocera:2014gqa, deFlorian:2014yva}.\par

Similar measurements for dijet production are also made in order to better constrain the shape of the gluon helicity. Right panel of Fig.~\ref{fig-1} presents the impact of all the RHIC results on gluon polarization from a most recent global analysis by DSSV group. It shows the integral of the gluon helicity within two distinct momentum fraction regions. The horizontal axis delineates the region from $0.05 < x < 1$, while the vertical axis is $0.001 < x < 0.05$. Significant polarization of gluon has been observed inside proton, with $\sim8 \sigma$ significance over zero in $0.05 < x < 1$ (0.218 $\pm$ 0.027 at 68\% C.L.)~\cite{RHICSPIN:2023zxx}.\par 

\section{Three Dimensional Tomography of the Nucleon: TMD}
\label{sec-2}

\begin{figure*}
\centering
\includegraphics[width=13cm,clip]{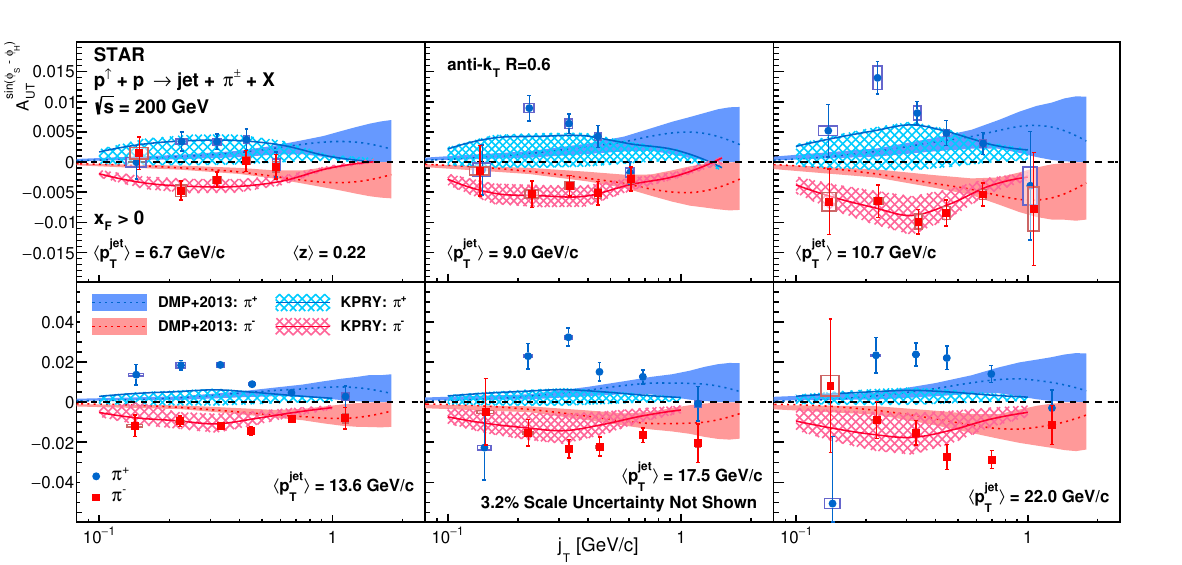}
\caption{Collins asymmetries as a function of the charged pion’s momentum transverse to the jet axis, $j_{T}$, in six different jet $p_{T}$ bins from~\cite{STAR:2022hqg}.}
\label{fig-2}       
\end{figure*}

To completely understand the proton’s internal spin structure, we need to go beyond the one-dimensional collinear PDF, and pursue the three-dimensional tomography of the nucleon. The transverse momentum dependent physics (TMDs) seeks to elucidate the correlations between transverse momentum and the intrinsic spin of quarks and gluons within a proton. This framework offers a three-dimensional tomographic perspective on the internal structure of hadrons, revealing nuanced distributions of partons in both momentum and spin~\cite{Boussarie:2023izj}.\par

The distribution of an unpolarized hadron inside the jet, which is initiated by a transversely polarized quark, provides a novel way to probe the Collins fragmentation function coupled with quark transversity. The correlation between the transverse spin of the quark and the transverse momentum of the fragmenting hadron relative to the jet axis, denoted as $j_{T}$, leads to a distinctive azimuthal modulation in the hadron distribution, also called Collins asymmetry. This phenomenon provides valuable insights into the dynamics of transverse spin and momentum within the context of jet formation in transversely polarized $pp$ collisions~\cite{Yuan:2007nd, DAlesio:2010sag}.\par

Figure~\ref{fig-2} shows the Collins asymmetries for charge pions within jets in polarized $pp$ collision at $\sqrt{s} =$ 200 GeV that STAR published last year~\cite{STAR:2022hqg}. The results are presented using a multidimensional binning across six distinct jet $p_{T}$ bins with integrated luminosity of 74~pb$^{-1}$ from 2012 and 2015 running periods. This helps to elucidate the interplay of parton's transverse motion in both momentum and spin dependence from TMD physics. The theoretical predictions are based on the DMP+2013~\cite{DAlesio:2017bvu} and KPRY~\cite{Kang:2017btw} models, combining quark transversity from Semi-Inclusive Deep Inelastic Scattering with the Collins fragmentation function from $e^{+}e^{-}$ annihilation, under the assumption of universality and factorization in $pp$ collisions. However, notable discrepancies are observed between the data and both model calculations, indicating a need for further theoretical efforts to understand these differences. Additionally, in the upcoming 2024 running period, twice more luminosity data from transversely polarized $pp$ collisions at $\sqrt{s} =$ 200 GeV will be recorded, which will further improve the precision of this measurement.\par

\section{Summary}

Various spin physics results from STAR have made significant progress in advancing our understanding of the internal spin structure of nucleons. STAR has completed the longitudinal polarized data taken, with few remaining results to be published soon. Notably, evidence has been found for positive gluon polarization inside proton in the $x > 0.05$ region. The venture into transverse spin measurements has produced impactful findings, including the observation of the transversity, Sivers, and Collins effects in proton-proton collisions. With recent upgrades, STAR will continue to unravel the complexities of nucleon spin dynamics with its unique kinematic coverage at both mid and forward rapidity before the EIC.\par

\section{Acknowledgment}
This work was supported in part by the National Natural Science Foundation of China under Grant No.12375139 and the Qilu Youth Scholar Funding of Shandong University.\par

%
%
%

\end{document}